\documentclass[aps,twocolumn,floatfix,amssymb,amsmath,tightenlines,prb]{revtex4}
\usepackage{graphicx}
\usepackage{bm}
\usepackage{dcolumn}

\newcommand{\deff}{\ensuremath{D_\mathrm{eff}}}
\newcommand{\A}{\ensuremath{\mathrm{\mathring{A}}}}
\newcommand{\Asqps}{\ensuremath{\mathrm{\mathring{A}}^2/\mathrm{ps}}}
\newcommand{\scap}{\scriptscriptstyle}

\begin{document}

\preprint{laur-02-4552}

\title{Non-Gaussian dynamics from a simulation of a short peptide:\\
Loop closure rates and effective diffusion coefficients}

\author{John J. Portman}
\affiliation{Center for Nonlinear Studies and Theoretical Division, 
Los Alamos National Laboratory, Los Alamos, New Mexico 87501}

\date{October 31, 2002; laur-02-4552}

\begin{widetext}
\begin{abstract}
Intrachain contact formation rates, fundamental to the dynamics 
of biopolymer self-organization such as
protein folding, can be monitored in the laboratory through fluorescence
quenching measurements.  
The common approximations for the intrachain contact rate given by 
the theory of 
Szabo, Schulten, and Schulten (SSS)
[J.\ Chem.\ Phys. {\bf 72}, 4350 (1980)] 
and Wilemski--Fixman (WF) 
[J.\ Chem.\ Phys. {\bf 60},878 (1973)] 
are shown to be complementary variational bounds:
the SSS and WF approximations are
lower and upper bounds, respectively, on the mean first contact times.
As reported in the literature,  the SSS approximation
requires an effective diffusion coefficient 10 to 100 times
smaller than expected
to fit experimentally measured quenching rates.
An all atom  molecular dynamics simulation of an eleven residue 
peptide sequence in explicit water is analyzed to investigate the 
source of this surprising parameter value.
The simulated diffusion limited contact time is approximately 
$6\,\mathrm{ns}$ for a reaction radius of
$4 \, \A$ for solvent viscosity corresponding to that of 
water at $298 \, \mathrm{K}$ 
and $1 \, \mathrm{atm}$
($\eta = 1.0 \, \mathrm{cP}$).
In analytical work, the polymer is typically  modeled by a Gaussian chain 
of effective monomers.
Compared to Gaussian dynamics, the simulated end-to-end distance
autocorrelation has a much slower relaxation.
The long time behavior of the distance autocorrelation function can 
be approximated by a Gaussian model
in which the monomer diffusion coefficient $D_0$ is reduced to $D_0/6$.
This value of the diffusion coefficient brings the mean end-to-end contact time
from analytical approximations and simulation into agreement 
in the sense that the SSS and WF approximations
bracket the simulated mean first contact time.
Attention to the non-Gaussian nature of the dynamics has direct implications
for the development of improved analytical models. 
\end{abstract}
\end{widetext}

\maketitle
 
\section*{Introduction}
Intrachain contact formation is fundamental to our understanding of 
biopolymer dynamics involved in  molecular self-organization.
Motivated primarily by the relevance to protein folding,
several groups have recently  measured the 
quenching rates of the terminal residues on short peptides
using fluorescent probes with short range 
quenching.\cite{ljl:wae:jh:99,bwhsdk:99,hhgn:02}
In these experiments, 
the fluorescence decay rate characterizes the rate of 
contact formation in peptides for which the dynamics should be 
less complicated (and hence, easier to interpret) 
than the folding dynamics of natural protein sequences.
Lapidus \emph{et al.}\cite{ljl:wae:jh:99} use a simple polymer
model with a single dynamical variable
to understand the measured rates, finding that the effective
diffusion constant needed to fit the data is 
approximately 30 times smaller than
the expected relative coefficient for free diffusion of the two 
terminal residues
(assuming this to be $1.5 \times 10^{-5} \,\mathrm{cm^2/s}$).
A small effective diffusion coefficient appears to be a rather
general result, consistent over different molecules and types
of probes.
Many years ago, Haas \emph{et al.}\cite{eh:ek:izs:78} 
also reported surprisingly small diffusion coefficients 
using similar polymer models
(see also Ref.~\onlinecite{haas:96}), 
but here the quenching mechanism of the terminal monomers of the peptides
is through the longer range fluorescence resonance energy transfer (FRET).
Recently, Wallace \emph{et. al}\cite{wybk:01} applied this analysis to
FRET quenching experiments in DNA, concluding that the effective
monomer diffusion coefficient was 1000 times slower than reported for
peptides.

Interpretation of this small effective diffusion coefficient
is particularly relevant to modern protein folding research.
The energy landscape theory of protein folding\cite{bosw:95} along
with advances in experimental techniques that probe faster timescales
has encouraged some researchers to focus on proteins with the
simplest kinetics, the so-called fast folding proteins.\cite{jackson:98,gruebele:99}
There has been considerable progress in understanding
these fast folding proteins from both experimental and
theoretical approaches.  For this group of proteins, 
the transition state ensemble probed experimentally through mutational 
studies pioneered by Fersht\cite{lsi:deo:arf:95} can be modeled fairly
accurately by simple polymer based models with interactions between 
distant residues specified by the native state 
topology.\cite{jjp:st:pgw:98,ovg:avf:99,ea:db:99,vm:wae:99,bas:jw:pgw:99}
Success of modeling the barrier crossing dynamics that determine folding
rates is far less clear; 
the few theories of the folding dynamics proposed so 
far\cite{dad:wag:99,jjp:st:pgw:01b,munoz:01,mkpm:02} 
could greatly benefit from more experimentally supported
microscopic parameters characterizing the peptide dynamics.
Such fundamental timescales and model parameters can be obtained
in principle from measured intrachain quenching rates and more 
accurate theoretical modeling.

Effective parameters are relevant to specific models.
A common model for polymer dynamics approximates the polymer backbone
connectivity by harmonic bonds between effective 
monomers.  As emphasized by Zwanzig,\cite{zwanzig:74}
a general harmonic potential energy can be defined through the static
correlations between the monomers.  
Interactions between distant  monomers
do not enter explicitly into the model except possibly (and indirectly)
through the static correlations defining the Gaussian potential.
The dynamics of these phantom chain models
obey Gaussian statistics. In particular, the dynamics of the 
end-to-end vector $\bm{R}(t)$ is
completely described by the Gaussian  Green's function (assuming
isotropy) 
\begin{widetext}
\begin{equation}\label{eq:G_R}
G(\bm{R}t|\bm{R}_0) = 
\left(\frac{3}{2\pi\left\langle R^2 \right\rangle[1 - \phi^2(t)]}
\right)^{3/2} 
      \exp \left[ 
	- \frac{3}{2\left\langle R^2 \right\rangle}
          \frac{ \left|\bm{R} - \phi(t)\bm{R}_0\right|^2}
                     {1 - \phi^2(t)}
           \right].
\end{equation}
\end{widetext}
Here, $\phi(t)$ is the equilibrium vector correlation
\begin{equation}
\phi(t) = \frac{\langle \bm{R}(t)\cdot \bm{R}(0)\rangle}
  {\left\langle R^2\right\rangle},
\end{equation}
and $\langle R^2 \rangle$ is the equilibrium mean square end-to-end
distance that characterizes the size of the polymer through the
Gaussian equilibrium distribution
\begin{equation}\label{eq:peq_R}
P_{\mathrm{eq}}(\bm{R}) 
= \left( \frac{3}{2 \pi\left\langle R^2 \right\rangle}\right)^{3/2} 
          \!\!\!  \exp \left[ -\beta U(\bm{R})\right],
\quad
\beta U(\bm{R}) = \frac{3R^2}{2\left\langle R^2 \right\rangle}.
\end{equation}

As long as the Gaussian nature of the dynamics is retained, the only
relevant difference between different Gaussian polymer models
is through the static size $\langle R^2 \rangle$ and the equilibrium
pair correlation $\phi(t)$.  
Neglecting friction with memory,\cite{zwanzig:74,mb:rz:78} 
$\phi(t)$ is typically expressed
as a sum of exponentials with a wide distribution of decay times.
Another common approximation treats $\bm{R}(t)$ as
the only dynamical variable, replacing $\phi(t)$ in Eq.(\ref{eq:G_R})
with a single exponential.  This single variable approximation to the
multi-bead dynamics is the one that has been used to analyze 
the fluorescence quenching experiments.

Some explanations for the
small effective diffusion coefficient required to fit measured quenching
rates can be incorporated into Gaussian models.
For example, Haas \emph{et al.}\cite{eh:ek:izs:78} 
suggested that the 
reduced effective diffusion constant may be  evidence for internal
friction of the chain (independent of the solvent viscosity).
Internal friction designed to capture the timescale of dihedral angle 
basin hopping has been introduced through a modified friction matrix 
while retaining the Gaussian description of the dynamics.\cite{cwm:mcw:85,fixman:89}
Alternatively, Lapidus \emph{et al.}\cite{ljl:wae:jh:99} 
noted that the reduced effective diffusion constant is reminiscent of 
diffusion in a one-dimensional
rugged potential well as discussed by Zwanzig.\cite{zwanzig:89}
Within the single variable approximation,
averaging over the rugged energy potential can lead to dynamics described
by diffusion on a smooth potential with a reduced effective diffusion constant.
Both local chain dynamics or energetic ruggedness can provide additional
friction that effectively slows the dynamics of the Gaussian chain.
A main focus of this paper is to determine whether a Gaussian chain
with modified parameters can capture timescale of the polymer dynamics
suggested by the small effective diffusion coefficient.

There are certainly other mechanisms that could slow the dynamics,
some even leading to glassy behavior.
The ruggedness in Zwanzig's one dimensional model are local extrema
in the potential along the reaction coordinate, sometimes referred to as
``longitudinal ruggedness.''  Alternatively, the Bryngelson and Wolynes theory 
of protein folding dynamics\cite{jdb:pgw:89} accounts for 
``transverse ruggedness'' (the coupling to hidden or additional degrees 
of freedom) that may induce kinetic traps along the reaction coordinate.
Interactions between monomers along the polymer chain can also generate a
form of friction. 
Mode-coupling theories predict that the dynamics of random heteropolymers
may become non-ergodic,\cite{dt:va:jkb:96,st:jjp:pgw:97} similar to spin 
glass dynamics below the dynamical glass transition 
temperature.\cite{trk:dt:87} 
Although one would not expect a short peptide to exhibit glassy behavior,
similar mechanisms may still be relevant to some degree in this case as well.

The general problem of intrachain quenching is a difficult
one that has not been solved explicitly even for 
a Gaussian chain.\cite{fn:jjp02_sokolov}
Two widely used approximations are the closure approximation proposed by
Wilemski and Fixman\cite{gw:mf:73b} (WF)  and 
the single variable approximation solved by 
Szabo, Schulten, and Schulten\cite{as:ks:zs:80} (SSS).
In this paper, it is shown that the diffusion limited
reaction rate from these two approximations are related through a variational
expression: the WF and SSS approximations are upper and lower bounds, 
respectively, to the mean first passage time of end-to-end contact of the
original chain.
Since the single variable approximation used to interpret the experimental
measurements is a lower bound, it is possible that the SSS approximation
requires a small effective diffusion constant to fit the experimental 
quenching time because it is a poor approximation to the exact 
mean contact time of the multi-bead chain.
Similarly, the WF approximation (had it been used) requires a larger
diffusion constant to fit the data, since it gives a longer contact
time than the SSS approximation.

The accuracy of these approximations have been assessed by the
comparison  to simulations of a Rouse chain 
(monomers sequentially connected by harmonic springs).
\cite{rwp:rz:as:96,gs:ay:bb:01,avb:ks:mt:2002}
The results reported by Pastor \emph{et al.}\cite{rwp:rz:as:96}
show that the SSS approximation underestimates the Rouse chain mean contact
time by roughly a factor of three
(although this factor is just an example as it depends
strongly on chain length and contact distance). 
Consequently, the accuracy of the lower bound may at least account for 
part of the small effective diffusion constant that needs clarification.

In order to develop improved models informed by measured quenching
rates, it is crucial to clarify the way in which the effective diffusion 
coefficient reflects the timescale of the polymer dynamics.
This question is addressed in this paper through an analysis
molecular dynamics simulation of a realistic peptide.  
To connect with analytic polymer models,
a general Gaussian chain is chosen to define the effective diffusion 
coefficient and compare against the simulated dynamics.

The peptide sequence $\mathrm{C(AGQ)}_3\mathrm{W}$  
(denoted by $\mathrm{CW}_3$)
studied experimentally by Lapidus \emph{et al.}\cite{ljl:wae:jh:99}
is simulated using all atom molecular dynamics in explicit solvent.
A closely related work by Yeh and Hummer\cite{iy:gh:02}
was recently published that
presents extensive all atom simulations of the shorter peptides
$\mathrm{C(AGQ)}_1\mathrm{W}$ and $\mathrm{C(AGQ)}_2\mathrm{W}$.
The simulated mean quenching time for $\mathrm{CW}_3$ is the same
order of magnitude as found previously for the shorter peptides:\cite{iy:gh:02}
for diffusion limited quenching with a reaction radius of $4 \, \A$ we
find mean contact time of approximately $6 \, \mathrm{ns}$ 
(adjusted to solvent viscosity $\eta^{\mathrm{wat}} = 1.0 \, \mathrm{cP}$).
The experimentally measured quenching time for this 
peptide (on the order of $100 \, \mathrm{ns}$)
is many times longer than the 
diffusion limited contact time,\cite{ljl:wae:jh:99}
suggesting that the quenching rate for this system is reaction limited not 
diffusion limited.
Indeed, the analysis of Yeh and Hummer\cite{iy:gh:02}
shows that the simulation is consistent with experiment assuming short 
distance quenching with a finite rate.\cite{ljl:wae:jh:01}
While this is very important for interpreting these particular experiments,
the fundamental time scale that one wishes to ultimately
understand in the folding context is the polymer dynamics of 
intrachain contacts, not the quenching properties of the probes.
The simulation of $\mathrm{CW}_3$ presented in this paper as well as 
those presented by Yeh and Hummer\cite{iy:gh:02} show that 
a reduced effective diffusion coefficient is still
required to fit even the shorter diffusion limited quenching time.
This is the focus of the present paper as the reaction limited quenching 
is a rather separate issue.

The analysis of the present simulation suggests that the small effective
diffusion coefficient does not originate from
local internal friction or longitudinal ruggedness that can still be modeled 
in an otherwise Gaussian formalism. 
The accuracy of the single variable approximation to the 
multi-bead quenching time does not seem essential to the discrepancy
between calculated and measured rates either.
For this simple reaction, the {\it prima facie} reaction 
coordinate of the quenching would be expected to be 
the end-to-end distance $R(t) = |\bm{R}(t)|$.
Consequently, $R(t)$ is also the relevant quantity requiring
an accurate dynamical description.
Due to Eq.(\ref{eq:G_R}),
the pair correlation $\langle R(t)R(0) \rangle$ for a Gaussian
model can be found once $\langle R^2 \rangle$ and $\phi(t)$ are specified.   
From the simulations, it is found that 
although $P_{\mathrm{eq}}(R)$ is roughly 
Gaussian and $\phi(t)$ is well described by superimposed
modes, $\langle R(t) R(0) \rangle$ relaxes much more slowly than
the Gaussian expression.
This is independent of a specific polymer model, relying
only on an assumption of isotropic Gaussian dynamics.
The simulated $\langle R(t)R(0) \rangle$ can
be approximately fit for long times with Gaussian dynamics
by reducing the diffusion coefficient to a value that 
also brings the simulations into agreement
with the bounds on the mean contact time
provided by the WF and SSS approximations.

The most fruitful extensions to the 
Gaussian formalism would likely include an additional
time scale or reaction coordinate that influences the end-to-end dynamics,
rather than improving the description 
of $\phi(t)$ through more elaborate Gaussian chain models.  
Furthermore, since the contact rate is directly related to the autocorrelation
$\langle R(t) R(0) \rangle$,
it may be more convenient theoretically to focus on this correlation than
the intrachain mean first passage time to improve polymer dynamics models.

\section*{Gaussian Chains}
Although the conclusions (and the analysis of the simulations)
do not rely on a specific Gaussian polymer dynamics model,
introducing a rather general Gaussian chain at the outset makes
the formalism more clear.  
Consider a chain consisting of $N$ monomers
with positions $\{\bm{r}_i\}$ and isotropic Gaussian connectivity
\begin{equation}\label{eq:U_chain}
\beta U[\{\bm{r}\}] = \frac{3}{2b^2} \sum_{ij} 
      \bm{r}_i \cdot \Gamma_{ij} \cdot \bm{r}_j,
\end{equation}
where $\beta = 1/k_B T$ is the inverse temperature, and
$b$ is the root mean square distance between adjacent monomers 
along the chain.  
The static correlations of monomer positions are determined by
the coefficients $\Gamma_{ij}$:
$\langle \bm{r}_i\cdot\bm{r}_j \rangle/b^2 = \Gamma_{ij}^{-1}$,
where $\langle \cdots \rangle$ denotes an equilibrium average over
the equilibrium distribution 
$P_{\mathrm{eq}}[\{\bm{r}\}] \sim \exp [-\beta U[\{\bm{r}\}]$.
(In this paper, spatial isotropy of matrices is implicit, e.g.,
$\Gamma_{ij}$ is the direct product of an $N \times N$ matrix
and the $3 \times 3$ indentity matrix.)

The dynamics are assumed to be
over-damped and Markovian with a spatially independent friction matrix 
$\gamma_{ij}$.
The probability distribution $P[\{\bm{r}\},t]$ evolves according
to the Smoluchowski equation 
\begin{equation}\label{eq:smol_chain}
\partial_t P[\{\bm{r}\},t] = {\cal D}[\{\bm{r}\}] P[\{\bm{r}\},t],
\end{equation}
where the diffusion operator is defined as
\begin{equation}\label{eq:diff_op_chain}
{\cal D}[\{\bm{r}\}] = 
\sum_{ij} \frac{\partial}{\partial \bm{r}_i} P_{\mathrm{eq}}[\{\bm{r}\}]
          \cdot D_{ij} \cdot
          \frac{\partial}{\partial \bm{r}_j} 
           \frac{1}{P_{\mathrm{eq}}[\{\bm{r}\}]}
\end{equation}
with the diffusion matrix $D_{ij} = k_B T \gamma_{ij}^{-1}$.

For a Gaussian chain, the equilibrium correlation functions and 
averages for any functions $f(\bm{R})$ and $g(\bm{R})$ of the end-to-end vector
$\bm{R} = \bm{r}_N - \bm{r}_i$
can be calculated as
\begin{equation}\label{eq:fgcorr_R}
\langle f(t)g(0) \rangle =
      \int \!\!\mathrm{d}^3\bm{R}
      \!\! \int \!\!\mathrm{d}^3\bm{R}_0 \,
            f(\bm{R})g(\bm{R}_0) G(\bm{R}t|\bm{R}_0) P_{\mathrm{eq}}(\bm{R}_0)
\end{equation}
\begin{equation}\label{eq:fave_R}
\langle f \rangle = 
 \int \!\! \mathrm{d}^3\bm{R} \, 
     f(\bm{R}) P_{\mathrm{eq}}(\bm{R}),
\end{equation}
where the 
Green's function $G(\bm{R}t|\bm{R}_0)$ 
and the equilibrium distribution $P_{\mathrm{eq}}(\bm{R})$ 
are given by Eq.(\ref{eq:G_R}) and Eq.(\ref{eq:peq_R}), respectively. 
Note that the chain dynamics are included through the dependence 
of $G(\bm{R}t|{\bm R}_0)$ on the pair correlation function $\phi(t)$.  
The form
of $\phi(t)$ for this general Gaussian chain is given the
Appendix.

End-to-end quenching is incorporated into the 
Smoluchowski equation as a distance dependent sink term
\begin{equation}\label{eq:smol_chain_sink}
\partial_t P[\{\bm{r}\},t] = {\cal D}[\{\bm{r}\}] P[\{\bm{r}\},t] 
  - \epsilon k(\bm{R}) P[\{\bm{r}\},t],
\end{equation}
where $\epsilon k(\bm{R})$ is the quenching rate for the relative 
end-to-end separation $\bm{R}$.  
Following Pastor \emph{et al.},\cite{rwp:rz:as:96}
I consider a Heavyside sink
\begin{equation}\label{eq:k_theta}
k(\bm{R}) = \begin{cases}
	            1 &,\quad |\bm{R}| \le a \\*
		    0 &,\quad |\bm{R}| > a
            \end{cases}
\end{equation}
with diffusion limited quenching, $\epsilon \rightarrow \infty$.
The reaction rate can be characterized by the mean first passage
time
\begin{equation}\label{eq:tau}
\tau = \int_0^{\infty} \!\!\!\! \mathrm{d}t \, 
       S(t) = \hat{S}(\omega = 0),
\end{equation}
where 
$\hat{S}(\omega)$ is the Laplace transform of the survival probability
\begin{equation}\label{eq:survival}
S(t) = \int \!\! \mathrm{d}^3\{\bm{r}\} \, 
     P[\{ \bm{r}\},t],
\end{equation}
$\hat{S}(\omega) = \mathrm{LT}_{t \rightarrow \omega}[S(t)] \equiv 
\int_0^{\infty} \!\mathrm{d}t \, e^{-\omega t} S(t)$.

The Laplace transformed survival probability (and, hence,
mean first passage time) for the Gaussian chain
obey complementary variational bounds.  
To develop this relationship, it is
necessary to introduce the adjoint Smoluchowski operator, 
\begin{equation}\label{eq:adj_op}
{\cal L}[\{\bm{r}\}] = 
- \sum_{ij} \frac{1}{P_{\mathrm{eq}}[\{\bm{r}\}]}
          \frac{\partial}{\partial \bm{r}_i}P_{\mathrm{eq}}[\{\bm{r}\}]
          \cdot D_{ij} \cdot
          \frac{\partial}{\partial \bm{r}_j},
\end{equation}
which is defined through 
$ {\cal D}(P_{\mathrm{eq}}f) =  - P_{\mathrm{eq}} {\cal L} f$,
for any function $f[\{ \bm{r} \},t]$.
Writing the probability distribution as 
$P(t) = \rho(t) P_{\mathrm{eq}}$
and assuming equilibrium initial conditions, 
Eq.(\ref{eq:smol_chain_sink}) becomes
\begin{equation}\label{eq:adj_sink}
\left[ 
\omega + {\cal L}[\{\bm{r}\}] + \epsilon k(\bm{R})
\right] 
\hat{\rho}[\{\bm{r}\},\omega]
= 1
\end{equation}
where $\hat{\rho} = \mathrm{LT}_{t \rightarrow \omega}[\rho]$.  In this
notation the survival probability is
$\hat{S}(\omega) = \langle \hat{\rho} \rangle$.

The variational bounds on the survival probability derived
in Ref.~\onlinecite{jjp:pgw:99}
can be written in terms of trial functions $\varphi[\{\bm r \}]$ 
and $\xi[\{\bm r \}]$ 
\begin{equation}\label{eq:bnds}
F_{\omega}[\varphi] \le \hat{S}(\omega) \le K_{\omega}[\xi],
\end{equation}
where the lower bound is given by
\begin{equation}\label{eq:f_lower}
F_{\omega}[\varphi] = - \langle \varphi( \omega + {\cal L} + \epsilon k)\varphi \rangle 
+ 2\langle \varphi \rangle ,
\end{equation}
and the upper bound  is given by\cite{fn:jjp02_kupper}
\begin{equation}\label{eq:k_upper}
\frac{1}{K_{\omega}[\xi]} = 
\omega + \frac{\epsilon \langle \xi k \rangle^2}
           {\langle \xi^2 k \rangle 
            + \mathrm{LT}_{t \rightarrow \omega} 
              \epsilon  \langle \Delta \left[ \xi(t)k(t) \right]
                                 \Delta \left[ \xi(0)k(0) \right] \rangle} 
\end{equation}
with $\Delta[\xi(t) k(t)] = \xi(t) k(t) - \langle\xi k\rangle$.
The optimal trial functions 
$\delta F_{\omega}[\varphi^*] = 0$ and 
$\delta K_{\omega}[\xi^*] = 0$ satisfy
Eq.(\ref{eq:adj_sink}), with 
$\varphi^* = \xi^* = \hat{\rho}$ and 
$F_{\omega}[\varphi^*] = K_{\omega}[\xi^*] = \hat{S}(\omega) $.
Note that Eq.(\ref{eq:bnds}) evaluated at $\omega = 0$ are bounds 
on the mean first passage time $\tau = \hat{S}(0)$.

In practice, the bounds on 
$\hat{S}(\omega)$ and $\tau$ provided
by Eqs.(\ref{eq:bnds}--\ref{eq:k_upper}) are determined by the
choice of trial function to optimize.
The upper bound on $\tau$ 
was first derived by Doi\cite{doi:75a}  who noted that 
$K_{\omega = 0}[\xi]$
evaluated at the simplest trial function $\xi = 1$
gives the well known Wilemski-Fixman approximation.\cite{gw:mf:73b}
For the diffusion limited reaction 
($\epsilon \rightarrow \infty$), the WF approximation becomes
\begin{equation}\label{eq:wf_tau}
\tau_{\scap{\mathrm{WF}}} = \int_0^{\infty} \!\!\! \mathrm{d}t \,
  	\left( \frac{\langle k(t) k(0)\rangle}{\langle k\rangle^2} - 1 \right).
\end{equation}

Similarly, the single variable approximation (like SSS) can be seen to be
a lower bound on $\tau$.
Restricting the trial function to 
functions of the end-to-end relative vector,
$F_{\omega}[\varphi(\bm{R})]$
can be reduced to averages over $P_{\mathrm{eq}}(\bm{R})$ which
are denoted by $\langle \cdots \rangle_{\bm{\mathrm{R}}}$.  
This is immediately evident 
for each term in Eq.(\ref{eq:f_lower}) except
$\langle \varphi {\cal L} \varphi \rangle$ which reduces according
to
\begin{eqnarray}\label{eq:L_reduce}
\langle \varphi {\cal L} \varphi \rangle
&=&
\sum_{ij} 
\left\langle 
    \frac{\partial \varphi(\bm{R})}{\partial \bm{r}_i}
    \cdot D_{ij} \cdot
    \frac{\partial \varphi(\bm{R})}{\partial \bm{r}_j} 
\right\rangle
\nonumber \\*
&=&
\left\langle
  \frac{\partial \varphi(\bm{R})}{\partial \bm{R}}
  \cdot \deff \cdot
  \frac{\partial \varphi(\bm{R})}{\partial \bm{R}}
\right\rangle_{\bm{\mathrm{R}}}
\\*
&=&
\langle \varphi {\cal L}(\bm{R}) \varphi \rangle_{\bm{\mathrm{R}}}.
\nonumber
\end{eqnarray}
Thus, for this class of trial function the lower bound becomes
(for $\omega = 0$)
\begin{equation}\label{eq:f_lower_R}
F_{\omega = 0}[\varphi(\bm{R})] = 
-\langle \varphi( {\cal L}(\bm{R}) + \epsilon k) \varphi 
                                                 \rangle_{\bm{{\mathrm{R}}}}
+ 2 \langle \varphi \rangle_{\bm{\mathrm{R}}},
\end{equation}
with the single variable adjoint operator
\begin{equation}\label{eq:adj_op_R}
{\cal L}(\bm{R}) = 
- \frac{1}{P_{\mathrm{eq}}(\bm{R})}
   \frac{\partial}{\partial \bm{R}} P_{\mathrm{eq}}(\bm{R})
   \cdot \deff \cdot
   \frac{\partial}{\partial \bm{R}},
\end{equation}
and effective diffusion constant
\begin{equation}\label{eq:deff}
\deff = D_{NN} + D_{11} - D_{1N} - D_{N1}.
\end{equation}
Optimizing Eq.(\ref{eq:f_lower_R}) with respect to $\varphi(\bm{R})$ 
gives\cite{fn:jjp02_1dvar}
$\left[{\cal L}(\bm{R}) + \epsilon k \right] \varphi = 1$.
For diffusion limited reactions ($\epsilon \rightarrow \infty$)
and a Heavyside sink,
the sink term can be represented 
by the boundary condition: $\varphi(\bm{R}) = 0$, 
for $|\bm{R}| \le a$.  Thus, the starting equation 
introduced by Szabo, Schulten and Schulten\cite{as:ks:zs:80}
\begin{equation}\label{eq:sss_eq}
{\cal L}(\bm{R})\varphi = 1, 
\qquad
 \varphi(|\bm{R}|\le a) = 0,
\end{equation}
is an approximation corresponding to a lower bound for
the mean first passage time for intrachain
contacts. The mean first passage time 
from  Eq.(\ref{eq:sss_eq}) 
($\tau = \langle \varphi \rangle$) 
can be solved exactly giving the SSS approximation 
\cite{as:ks:zs:80}
\begin{equation}\label{eq:tau_sss_quad}
\tau_{\scap{\mathrm{SSS}}} 
= \frac{1}{\deff}\int_a^{\infty} \!\!\! \mathrm{d}x \, 
           \frac{1}{x^2 \tilde{P}_{\mathrm{eq}}(x)}
       \left[ \int_x^{\infty}\!\!\! \mathrm{d}y \,
              y^2 \tilde{P}_{\mathrm{eq}}(y) \right]^2,
\end{equation}
where $\tilde{P}_{\mathrm{eq}}(R) \sim \exp[-\beta U(R)]$ is normalized
over $a \le R \le \infty$, and 
$\deff$ is given by Eq.(\ref{eq:deff}).

Putting this together, the mean first contact time for the original
polymer is bounded by
\begin{equation}\label{eq:tau_bnds}
\tau_{\scap{\mathrm{SSS}}} \le \tau \le \tau_{\scap{\mathrm{WF}}}.
\end{equation}
Consider applying these bounds to a simulation of a peptide.
Eq.(\ref{eq:tau_bnds}) states that the simulated mean contact time
is bounded by the SSS approximation calculated with the simulated 
potential of mean force and the WF approximation calculated
from the simulated sink-sink correlation function.
By itself, this is not very helpful in connecting to
to microscopic parameters of a polymer model that can 
be generalized to other problems, e.g., polymer based models
of protein folding dynamics.
However, the bounds calculated from the same many-bead polymer model
are useful since the exact $\tau$ from the model is generally unknown.
A chain with Gaussian dynamics is chosen in this paper to define
the model and associated parameters that are compared with the simulated
contact time and end-to-end dynamics.

In this case, the two approximations can be simplified further.
For the SSS approximation,
with $\beta U(R) = 3R^2/2\langle R^2 \rangle$, the inner integral 
of Eq.(\ref{eq:tau_sss_quad}) can be evaluated giving\cite{rwp:rz:as:96}
\begin{equation}\label{eq:sss_tau}
\tau_{\scap{\mathrm{SSS}}}\deff = 
\frac{\left\langle R^2 \right\rangle}{6\Gamma[3/2,x_0]}
\int_{x_0}^\infty \!\!\! \mathrm{d}u \, 
    u^{-3/2} e^u  \Gamma(3/2,u)^2,
\end{equation}
where $\Gamma(a,x)$ is the incomplete gamma function, and 
\begin{equation}\label{eq:x0}
x_0 = \frac{3a^2}{2\left\langle R^2 \right\rangle}.
\end{equation}
In the WF approximation,
the sink-sink correlation function in Eq.(\ref{eq:wf_tau}) can be
calculated using Eq.(\ref{eq:fgcorr_R}).  
For the case of a Heavyside sink, 
Wilemski and Fixman approximated the pair correlation using a
``unbalanced Heavyside delta approximation''\cite{gw:mf:73b}
that can be written as
\begin{equation} \label{eq:kk/k_wf}
\frac{\langle k(t)k(0) \rangle}{\langle k \rangle^2}
\approx 
\frac{
\mathrm{erf}(\xi)  - (2/{\sqrt \pi}) \xi \exp(-\xi^2)}
{\mathrm{erf}({\sqrt x_0}) 
    - (2/{\sqrt \pi}){\sqrt x_0} \exp(-x_0)},
\end{equation}
with $x_0$ given by Eq.(\ref{eq:x0}), and 
\begin{equation}\label{eq:kk_wf_var}
\xi(t) = {\sqrt \frac{x_0}{1-\phi^2(t)} }.
\end{equation}
The approximations 
Eq.(\ref{eq:sss_tau}) and Eqs.(\ref{eq:kk/k_wf} - \ref{eq:kk_wf_var}) 
were among those compared with simulations of Gaussian chains by 
Pastor \emph{et al.};\cite{rwp:rz:as:96} this work
verified (though did not emphasize)
that the two approximations are bounds for the mean contact times.

Since the comparison of these approximations to measured (or simulated)
contact times is framed in terms of the relative value
of the diffusion coefficient, this parameter
warrants closer consideration.
As shown in the Appendix, a short time approximation to the diffusion coefficient also
leads to $\deff$ given by Eq.(\ref{eq:deff}).
For a free draining chain (diagonal friction), 
$\deff = D_N + D_1$ is the relative diffusion constant
of the two end monomers, which becomes $\deff = 2D_0$
if the friction is uniform ($D_{ij} = D_0\delta_{ij}$).
Alternatively,  Pastor \emph{et al.}\cite{rwp:rz:as:96} also
derived  $\deff = 2D_0$ for a
Rouse chain (nearest neighbor connectivity) with uniform friction
using a ``local equilibrium'' approximation, integrating
over the internal monomer positions in Eq.(\ref{eq:adj_sink});
even though the formal manipulations are similar to 
the variational derivation,
it is not evident from the ``local equilibrium'' approximation
that the SSS approximation with this $\deff$ is actually 
a lower bound.
The diffusion coefficient contained the WF approximation
follows more naturally from the model through the time dependence
of the pair correlation function $\phi(t)$.

The two approximations to $\tau$ given 
by Eq.(\ref{eq:wf_tau}) and Eq.(\ref{eq:tau_sss_quad}) suggest
(in slightly different ways) that 
the dynamics of the end-to-end distance $R(t) = |\bm{R}(t)|$
should be given careful attention.
In the SSS approximation, Eq.(\ref{eq:sss_eq}) describes the dynamics of
$R(t)$ in an isotropic
potential of mean force $\beta U(R)$, 
i.e., $R(t)$ is treated as a single reaction coordinate.
A single variable treatment should be fine to calculate 
rates as long as the effective diffusion constant is appropriate.
In the formalism presented here, $\deff$ arises from
reducing a more microscopic description to an effective one.  
On the other hand, without the advantage of a microscopic theory,
$\deff$ could be determined from
the dynamics of the reaction coordinate (obtained from a simulation, for
example).
Defined in this way,  $\deff$ is
truly an {\it effective} diffusion constant of the theory, 
not necessarily connected to microscopic parameters in a straight-forward
way
(see Ref.~\onlinecite{nds:jno:pgw:96}, for example).
Turning now to the WF approximation, 
the dynamics of $R(t)$ enters explicitly
through the sink-sink correlation function, since it is the 
autocorrelation of a function $R(t)$.

The pair correlation 
$\phi(t)$ has both a distance and angular component:
$\phi(t) = \langle R(t)R(0)\cos \theta(t) \rangle/\langle R^2 \rangle$,
where $\cos \theta(t)$ is the angle between $\bm{R}(t)$ and $\bm{R}(0)$.
Separating the distance and angular components, we
consider the correlation functions
\begin{equation}\label{eq:phi_dth}
\phi_{\mathrm{d}}(t) 
= \frac{\langle R(t) R(0) \rangle}{\left\langle R^2 \right\rangle},
\qquad
\phi_{\theta}(t) = \langle \cos \theta(t) \rangle.
\end{equation}
The Gaussian Green's function [Eq.(\ref{eq:G_R})] imposes
a specific dependence of $\phi_{\mathrm{d}}(t)$ and $\phi_{\theta}(t)$
on $\phi(t)$, just as it does on the sink-sink correlation function
given in Eq.(\ref{eq:kk/k_wf}).
Performing the integrals in Eq.(\ref{eq:fgcorr_R}), 
gives 
\begin{eqnarray}
\label{eq:phi_d_g}
\phi_{\mathrm{d}}^{\mathrm{g}}(t) 
&=&
    \frac{2}{3\pi}
	\left[
	\frac{1 + 2 \phi^2(t)}{\phi(t)} \arcsin \phi(t)
	+ 3 \sqrt{1 - \phi^2(t)}
	\right] , \\*
\label{eq:phi_th_g}
\phi_{\theta}^{\mathrm{g}}(t) 
&=&
     \frac{2}{\pi}
	\left[
	\frac{2 \phi^2(t)-1}{\phi^2(t)} \arcsin \phi(t)
	+ \frac{\sqrt{1 - \phi^2(t)} }{\phi(t)}
	\right].
\end{eqnarray}
(The superscript indicates that this is a Gaussian relationship.)
Not only are each correlation a function of $\phi(t)$, 
but these Gaussian expressions obviously also have specific 
relationships to each other, e.g., $3\dot{\phi}_{\mathrm{d}}^{\mathrm{g}}(t) 
= \dot{\phi}(t)\phi_{\theta}^{\mathrm{g}}(t)$.

Both $\phi_{\mathrm{d}}^{\mathrm{g}}(t)$ and $\phi_{\theta}^{\mathrm{g}}(t)$ 
are normalized to 
one at $t=0$.  
Expanding in powers of $\phi(t)$ gives to lowest order 
\begin{equation}\label{eq:phi_dth_longtime}
\phi_{\mathrm{d}}^{\mathrm{g}}(t) 
\sim \frac{8}{3\pi}\left( 1 + \frac{1}{6}\phi^2(t) \right),
\qquad
\phi_{\theta}^{\mathrm{g}} \sim \frac{8}{3\pi}\phi(t).
\end{equation}
Thus, the Gaussian correlation
$\phi_{\mathrm{d}}^{\mathrm{g}}(t)$ approaches the long time limit 
$\langle R \rangle^2/\langle R^2 \rangle = 8/3\pi$
at twice the rate that $\phi(t)$ decays,
and $\phi_{\theta}^{\mathrm{g}}(t)$ approaches zero at the same rate 
as $\phi(t)$.

\section*{Simulation}
In this section, a molecular dynamics simulation of a short peptide is presented
to investigate the end-to-end polymer dynamics and the 
diffusion limited quenching rate approximations
developed in the previous section.
Although simplified models can be very
helpful in determining the essential underlying physics responsible for
realistic polymer dynamics, the simulation reported here is of a
detailed model of the $\mathrm{CW}_3$ peptide studied experimentally by 
Lapidus \emph{et al.}\cite{ljl:wae:jh:99}
As emphasized by Yeh and Hummer,\cite{iy:gh:02}
simulations of a realistic model provide a stringent 
test for the molecular potentials when
compared directly with measured quenching rates.
From a more theoretical point of view, 
simulations obviously also make accessible quantities essential to the 
consistency of the theory (e.g., correlation functions)
that are not always easy to measure in the laboratory.
Since this work is motivated primarily by the unexplained
parameter values needed to fit experimentally measured rates,
the simulation presented in this section is an attempt
to capture the dynamics relevant to the experimental system.

The peptide
$\mathrm{Ace-C(AGQ)_3W-Nme}$ 
is modeled by
the force field of Cornell \emph{et al.}\cite{cornell_parm:95}
The {\footnotesize AMBER} 4.1\cite{amber41:95} suite of programs are used,
modified to include the generalized reaction field treatment
of electrostatics\cite{lrp:gh:aeg:94,gh:dms:mn:94} with a 
cutoff of $9\,\A$.  
The system, solvated with approximately 1700 TIP3P water molecules, 
was held
at constant temperature ($T = 300\,\mathrm{K}$) 
and pressure ($P = 1\,\mathrm{atm}$)
by Berendsen couplings,\cite{bpgdh:84} each with
a relaxation time of $0.1\,\mathrm{ps}$.  The integration step is 
$0.002 \, \mathrm{ps}$.
Non-bonded pair list was updated every 10 integration steps.
A single trajectory of $50 \, \mathrm{ns}$  was generated, starting from an
extended conformation, with conformational coordinates saved
every $0.5\,\mathrm{ps}$.

\begin{figure}[htb]
\includegraphics[width=3.3in,clip=true]{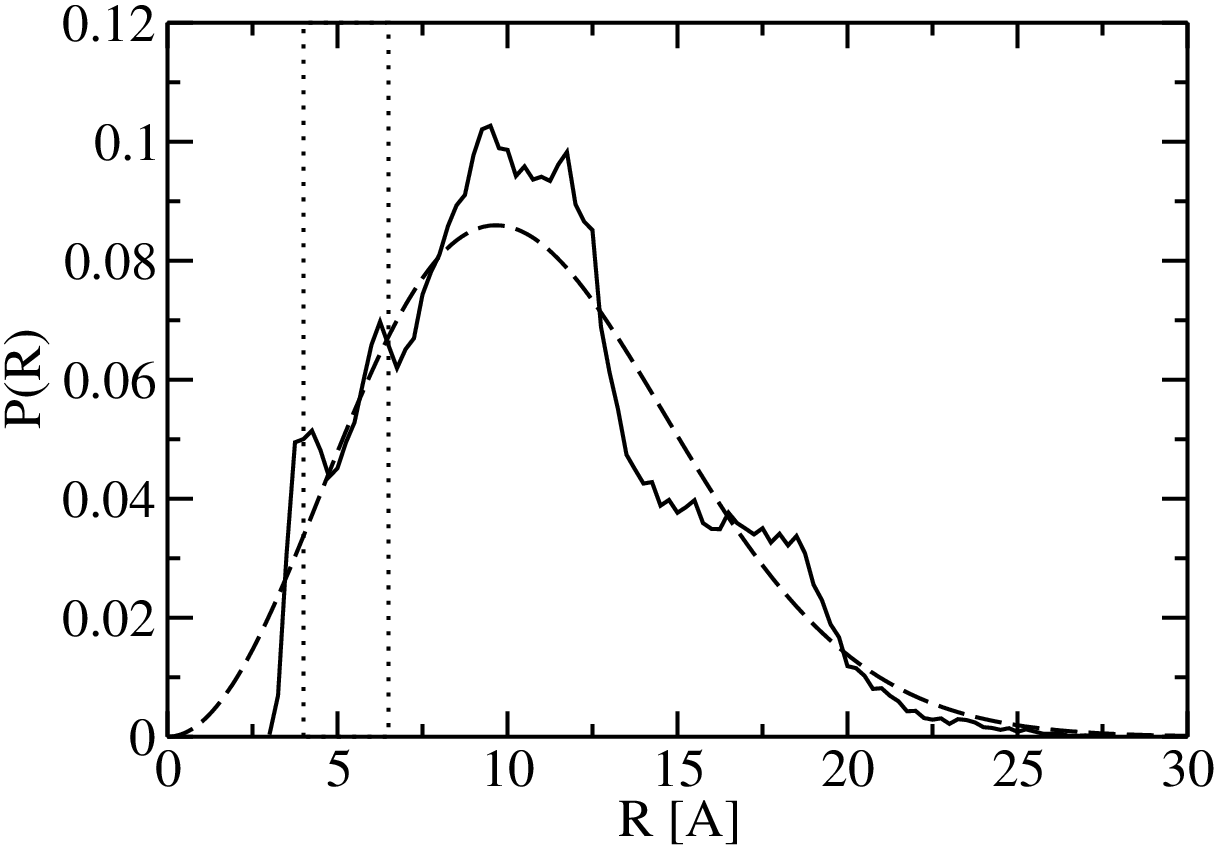}
  \caption{\label{fig:p_of_r} 
Probability distribution of the end-to-end distance for $\mathrm{CW}_3$. 
Solid: histogram from simulation with a bin width of 
$0.25\,\A$;
dashed: Gaussian distribution with simulated mean square distance 
$\langle R^2 \rangle = 140\,\A^2$.
The vertical dotted lines indicate the range of contact radii
$4 \, \A \le a \le 6.5 \, \A$.
}
\end{figure}

\begin{figure}[htb]
\includegraphics[width=3.3in,clip=true]{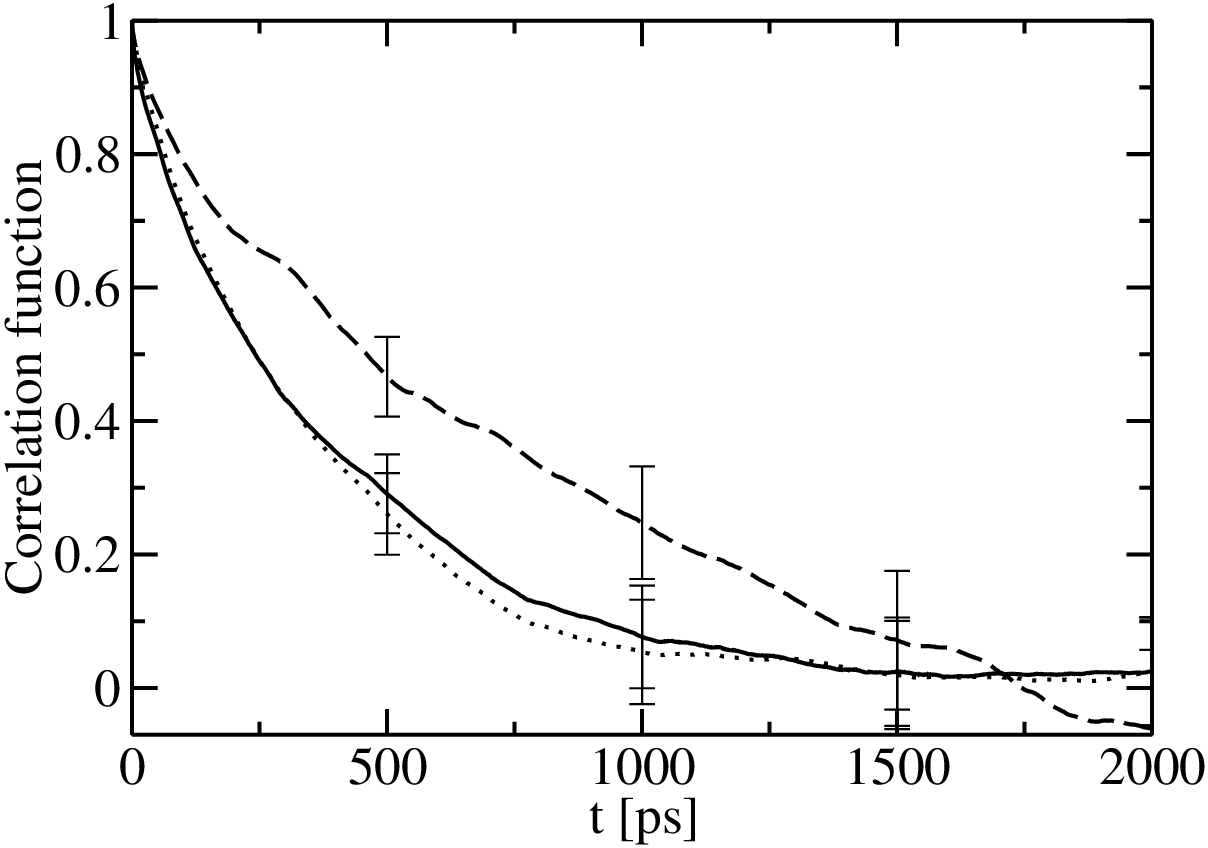}
  \caption{\label{fig:phi_d_th}
Simulated pair correlation functions of $\mathrm{CW}_3$.
Solid: $\phi(t)$;
dotted: $\phi_{\theta}(t)$;
dashed: normalized $\bar{\phi}_{\mathrm{d}}(t)$.
Error bars are computed according to  
Zwanzig and Ailawadi\protect\cite{rz:nka:69};
for example, the error bars for $\phi_{\theta}(t)$ are
$\pm \sqrt{2 T_{\theta}/T}(1 - \phi_\theta(t))$ 
where $T$ is total simulation time ($50\,\mathrm{ns}$), and 
$T_{\theta} = \int{\!\!\mathrm{d}}t \phi_\theta^2(t)$.
}
\end{figure}

The geometric center of all heavy atoms of the 
cystine and tryptophan sidechains (denoted by cm)
are chosen to define the end-to-end vector, 
$\bm{R}(t) = \bm{r}^{\mathrm{w}}_{\mathrm{cm}}(t) 
- \bm{r}^{\mathrm{c}}_{\mathrm{cm}}(t)$.  
Fig.~\ref{fig:p_of_r} shows
the probability distribution of the end-to-end distance
$R = |\bm{R}|$, as well as a Gaussian distribution
with the simulated mean square end-to-end distance,
$\langle R^2 \rangle_{\mathrm{sim}} \approx 140\,\A^2$.
Although there is some deviation, the distribution
can roughly be described as Gaussian.  The mean distance
$\langle R \rangle_{\mathrm{sim}}^2/\langle R^2 \rangle_{\mathrm{sim}}
\approx 0.87$ is within a few percent of the Gaussian
relationship
$\langle R \rangle^2/\langle R^2 \rangle
= 8/3\pi \approx 0.85$.

The simulated pair correlation functions, $\phi(t)$, 
$\phi_{\mathrm{d}}(t)$, and  $\phi_{\theta}(t)$
are shown in Fig.~\ref{fig:phi_d_th}.  
(Note the radial correlation function has been normalized:
$\bar{\phi}_{\mathrm{d}}(t) 
= \langle \delta R(t) \delta R(0) \rangle/\langle \delta R^2 \rangle$,
with
$\delta R(t) = R(t) - \langle R \rangle$.)
The pair correlation function
$\phi(t)$ follows the relaxation
of the rotational correlation function $\phi_{\theta}(t)$
in agreement with the limiting Gaussian expression
[Eq.(\ref{eq:phi_dth_longtime})].
The radial autocorrelation decays more slowly than
the others.  Although a more careful
evaluation is considered later, this slower relaxation 
can be seen to be a manifestation of non-Gaussian dynamics
of the end-to-end relative vector $\bm{R}(t)$
[\emph{c.f.}, Eq.(\ref{eq:phi_dth_longtime})].

\begin{figure}[htb]
\includegraphics[width=3.3in,clip=true]{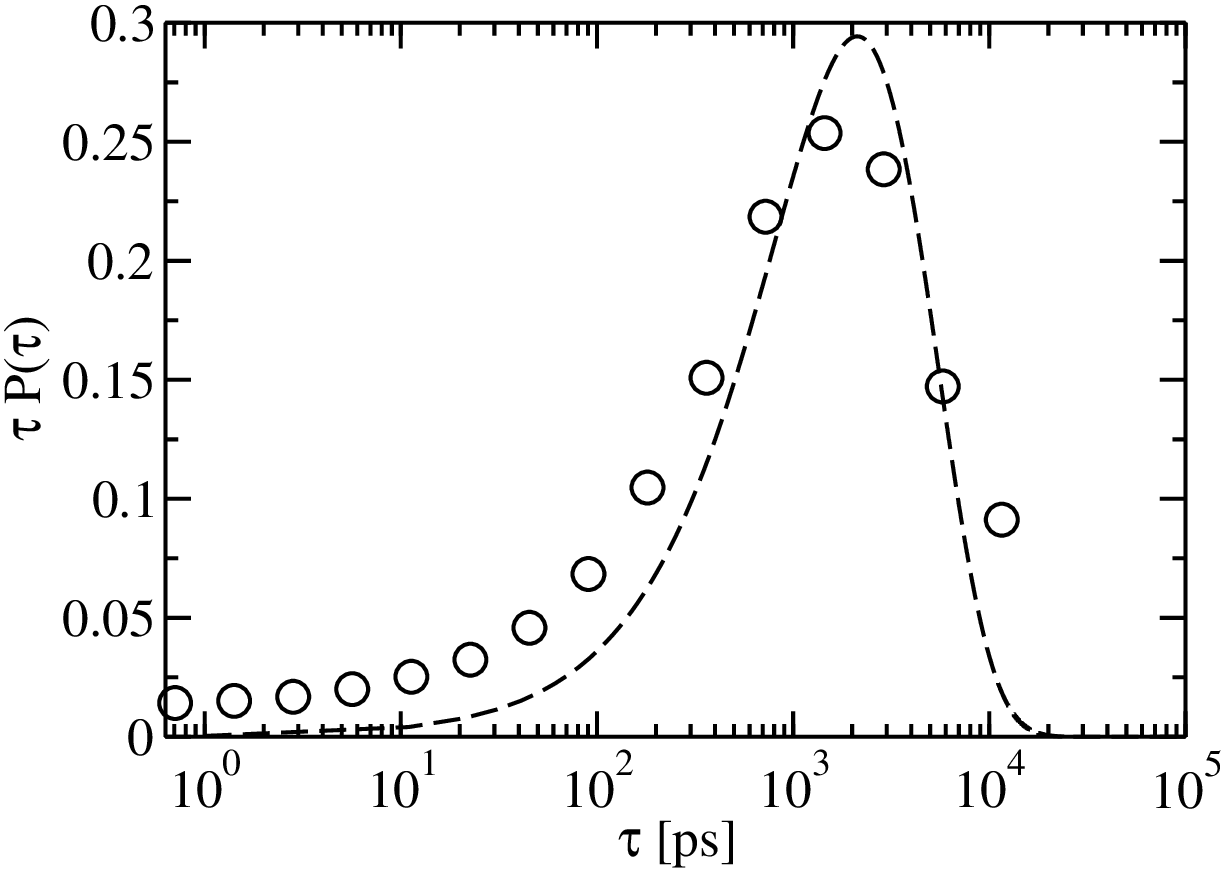}
  \caption{\label{fig:tau_dist}
First passage time distribution of $\mathrm{CW}_3$
for a Heavyside sink of radius $a = 5\,\A$.  
The histogram is on a
logarithmic scale with a bin width equal to powers of 2 (circles);
Dashed line: exponential distribution,
$P(\tau) = A \tau_{\mathrm{sim}}^{-1} \exp[-\tau/\tau_{\mathrm{sim}}]$, 
with
$\tau_{\mathrm{sim}} = \langle \Delta^2\rangle_{\mathrm{sim}}
/2\langle\Delta \rangle_{\mathrm{sim}} = 2123$ps,
and $A = 0.8$.
}
\end{figure}

The diffusion limited mean first passage time can be calculated
from the distribution of time intervals between 
contacts\cite{ljl:wae:jh:2002}
$\{\Delta_k\}$:
\begin{equation}\label{eq:delta}
\{ \Delta_k \}
= 
\{ t_j - t_i |
    R(t_{i-1})\le a;
    R(t_i)  > a; R(t_j) \le a;
 \},
\end{equation}
for $i < j$, and $k = 1 \ldots n_{\mathrm{int}}$.
Taking the initial distribution to be every conformation with
with $R(t) > a$ 
gives the equilibrium
distribution of contact times $P(\tau)$ with a mean contact time
$\tau_{\mathrm{sim}} = \langle \Delta^2 \rangle_{\mathrm{sim}}
/2\langle \Delta \rangle_{\mathrm{sim}}$.
The simulated mean first passage times of $\mathrm{CW}_3$ for 
different sink radii is given in Table~\ref{tbl:tau_cw3}.
For example, a sink radius of $a = 5.0\,\A$ had
$n_{\mathrm{int}} = 870$ separate contact events with a mean contact time 
$\tau_{\mathrm{sim}} = 2.1 \pm 0.7 \,\mathrm{ns}$.
As shown in Fig.~\ref{fig:tau_dist},
the distribution of $P(\tau)$  for a sink radius of $a = 5.0\,\A$
has a strong exponential component (seen by the width 
of a couple of decades) and a non-exponential component at short
times.  A non-exponential component is expected because
of short-time recrossings at the sharp edge defining the Heavyside
sink. 
Although the two components are not well separated,
the exponential component is described reasonably well by 
$P(\tau) = \tau_{\mathrm{sim}}^{-1}\exp(-\tau/\tau_{\mathrm{sim}})$ 
with an amplitude of 0.8.
The simulated distribution $P(\tau)$ for other sink radii are similar.
Note that the times reported are in the units from the simulation.  
Anticipating the discussion in the next section,  to approxmately
convert this time to units consistent with the viscosity of water at
$298 \, \mathrm{K}$ ($\eta^{\mathrm{wat}} = 1.0 \, \mathrm{cP}$),
$\tau_{\mathrm{sim}}$ 
should multiplied by a factor of two;
for example, the mean time to make a contact at a distance 
$a = 4.0\,\A$ corresponds to
$\tau_{\mathrm{sim}}^{\mathrm{wat}} = 5.6 \pm 1.4 \,\mathrm{ns}$.

\section*{Free Diffusion Coefficient}
A good estimate of the diffusion coefficient of the
end monomers in TIP3P water is needed to compare the 
simulated contact times with analytical models. 
To this end, two separate $10\,\mathrm{ns}$ 
simulations of $\mathrm{H_2-C-OH}$ and $\mathrm{H_2-W-OH}$ 
in approximately 1180 TIP3P waters were carried out.
Otherwise, the conditions are the same as those described for the peptide
simulation.
The value determined for the free diffusion
coefficient $D_0$ appropriate for the simulation will be used in the 
next section to compare with analytic models.

\begin{figure}[htb]
\includegraphics[width=3.3in,clip=true]{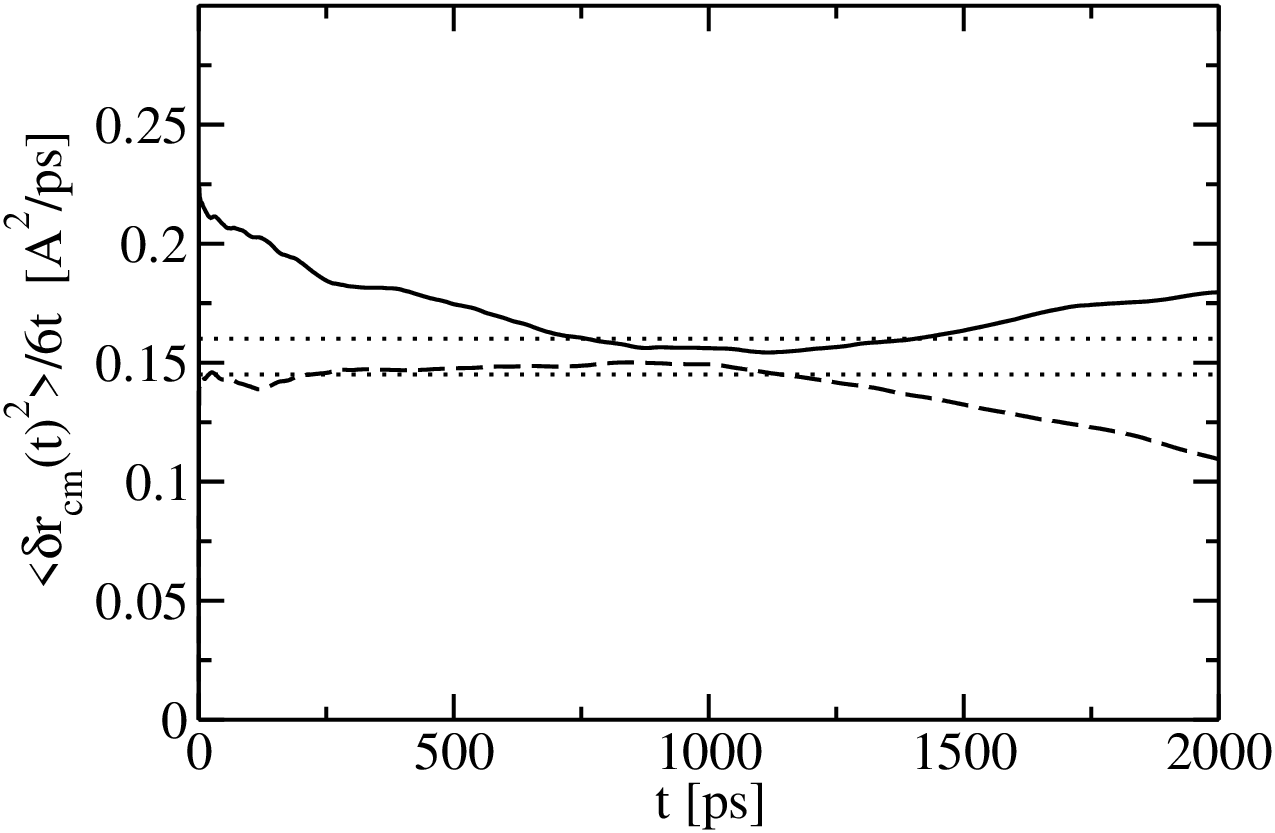}
  \caption{\label{fig:diff} 
Simulated diffusion coefficient for cystine (solid) and 
tryptophan (dashed) in TIP3P water.  
The dotted lines show the mean values over 
$500\,\mathrm{ps}$ to $1500\,\mathrm{ps}$,
$D_0^{\mathrm{c}} = 0.16\,\Asqps$
and  $D_0^{\mathrm{w}} = 0.145\,\Asqps$.
}	
\end{figure}

The translational diffusion coefficient of the geometric center of the 
heavy atoms of the residues 
can be calculated from the Einstein relation
$D \sim \langle \delta \bm{r}_{\mathrm{cm}}(t)^2 \rangle/6 t$ for long times, where
$\delta \bm{r}_{\mathrm{cm}}(t) 
       = \bm{r}_{\mathrm{cm}}(t) - \bm{r}_{\mathrm{cm}}(0)$.
The diffusion coefficient for cystine and tryptophan are shown 
in Fig.~\ref{fig:diff}.  Although there is still some variation due to the
finite time of the simulation, 
the average diffusion coefficient over 500ps to 1500ps is taken 
as an estimate, giving
$D_0^{\mathrm{c}} = 0.16 \, \Asqps$ for cystine and 
$D_0^{\mathrm{w}} = 0.145 \, \Asqps$ for tryptophan.  
We can verify these values using Stokes Law, $k_{\mathrm{B}}T/D_0 = 6\pi \eta_0 a_0$,
with $a_0$ equal to the van der Waals radii of the sidechains,\cite{creighton:93}
to find the viscosity of TIP3P water, $\eta_0$.
For cystine, using $a_0 = 2.75\,\A$ 
gives $\eta_0 = 0.50\,\mathrm{cP}$; and
for tryptophan, using $a_0 = 3.38\,\A$ 
gives $\eta_0 = 0.45\,\mathrm{cP}$;
the two estimates differing by 10\%.
This is in reasonable agreement with the viscosity of TIP3P calculated
by Shen and Freed, $\eta_0 = 0.506 \pm 0.043\,\mathrm{cP}$.\cite{ms:kff:02}

The effective diffusion constant for the single variable
(SSS) approximation of the mean first contact time is taken to be the
relative diffusion constant of cystine and tryptophan,
$D_\mathrm{eff} = D_0^\mathrm{c} + D_0^{\mathrm{w}} = 3.0 \times 10^{-5} \, \mathrm{cm^2/s}$.
Translating this to a uniform monomer friction we have
$D_0 = \deff/2 = 1.5 \times 10^{-5} \mathrm{cm^2/s}$.
This value for the effective diffusion constant is
the central assumption in the comparison of the simulation
with Gaussian chain models.

\begin{figure}[htb]
\includegraphics[width=3.3in,clip=true]{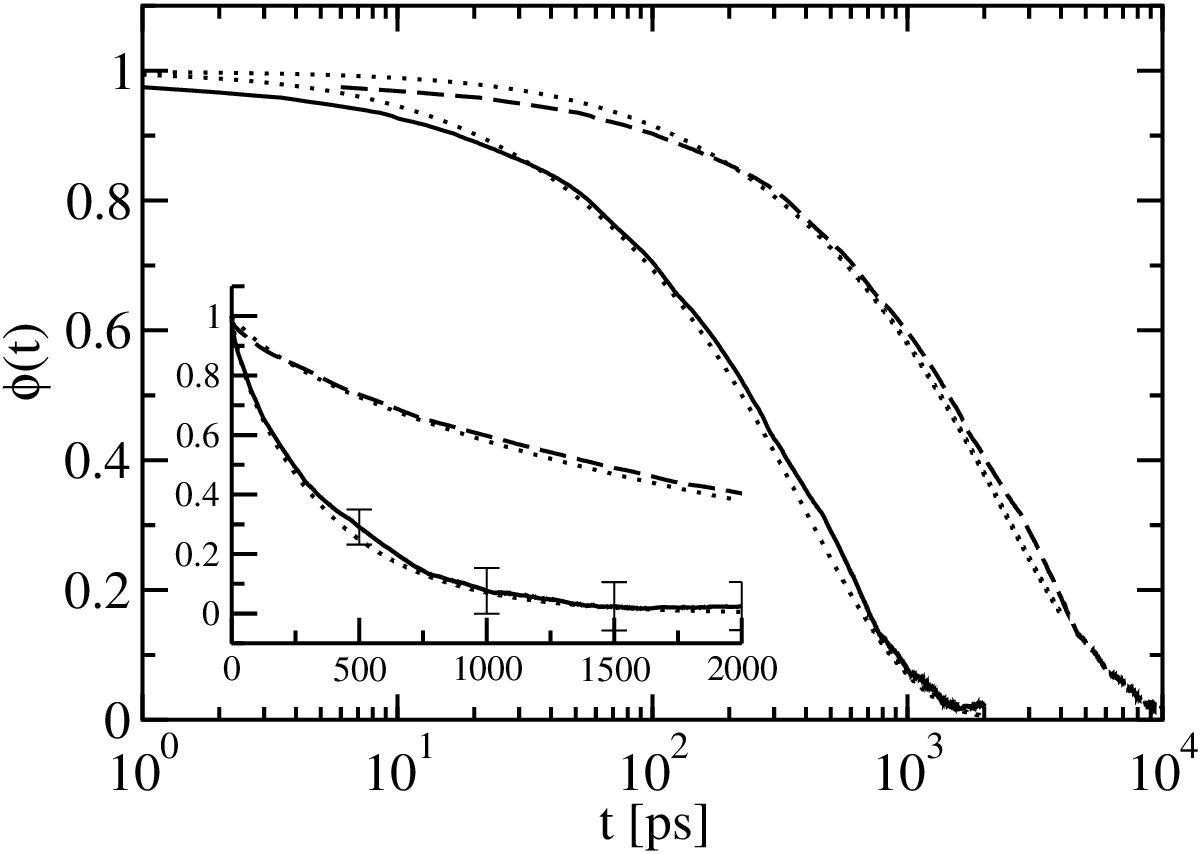}
  \caption{\label{fig:phi}
Pair correlation $\phi(t)$ for $\mathrm{CW}_3$. 
Lower curves:
$\phi(t)$ from simulation (solid), and Rouse chain (dotted)
with 
$N = 11$, $b = 3.8 \,\A$, and $D_{ij} = D_0\delta_{ij}$, 
$D_0 = 1.5 \times 10^{-5} \mathrm{cm^2/s}$.
Upper curves, scaled diffusion coefficient, $\bar{D}_0 = D_0/6$: 
$\phi(t)$ from simulation with time scaled
$t \rightarrow \bar{t} = (D_0/\bar{D}_0) t$ (dashed),
and Rouse chain [Eqs.(\ref{eq:rouse_modes}--\ref{eq:rouse_deff})]
with monomer diffusion coefficient $\bar{D}_0$ (dotted).
Inset is same plot on linear time axis.
}
\end{figure}

Identifying
$D_\mathrm{eff}$ as the relative diffusion
constant of the two end monomers is only rigorously 
true for free-draining chains
[Eq.(\ref{eq:deff})]. Comparison of the
pair correlation function $\phi(t)$ from simulation and 
a free-draining chain supports for this choice (see Fig.~\ref{fig:phi}).  
For times longer than
about $50\,\mathrm{ps}$, $\phi_{\mathrm{sim}}(t)$ agrees closely
with  $\phi(t)$ from a Rouse chain 
[Eqs.(\ref{eq:rouse_modes}--\ref{eq:rouse_deff})] with diagonal friction,
$D_{ij} = D_0\delta_{ij}$.  The Rouse chain model uses reasonable parameters
for this peptide, where the monomers of the chain are the $\alpha$-carbons
of the peptide: $N = 11$, $b = 3.8 \, \A$, and 
$D_0 = 1.5 \times 10^{-5} \mathrm{cm^2/s}$ is the 
monomer diffusion coefficient determined above.
The mean square distance of this model Rouse chain, 
$\langle R^2 \rangle = (N-1)b^2 = 144\,\A$,
agrees with the simulated value of 
$\langle R^2 \rangle_{\mathrm{sim}} = 140 \A$.
The close agreement of the static and dynamic correlations
is quite surprising,  
and one shouldn't expect it to hold generally, e.g., for all
chain lengths.  
The correspondence with a 
Rouse chain is considered only to support the choice of the effective
diffusion constant $D_{\mathrm{eff}} = 2 D_0$, with 
$D_0 = 1.5 \times 10^{-5} \mathrm{cm^2/s}$
---  the remaining analysis
assumes only this value for $D_{\mathrm{eff}}$ and
Gaussian statistics of $\bm{R}(t)$.

Except where noted, the reported values for the diffusion coefficient 
and contact time
correspond to the simulation viscosity $\eta_0 \approx 0.5 \, \mathrm{cP}$.  
To approximately convert to the viscosity of water at $298 \, \mathrm{K}$ 
and $1 \, \mathrm{atm}$, 
$\eta^{\mathrm{wat}} = 1.0 \, \mathrm{cP}$, 
the diffusion coefficient $D_0$ or contact time $\tau$ should be 
divided or multiplied by a factor of two, respectively.
For example, the relative diffusion coefficient corresponding to 
$\eta^{\mathrm{wat}} = 1.0 \, \mathrm{cP}$
is 
$2 D_0^{\mathrm{wat}} = 1.5 \times 10^{-5} \mathrm{cm^2/s}$.

\section*{Comparison of Simulation with Gaussian Dynamics}
In this section, the simulated mean contact time and correlation
functions of $\mathrm{CW}_3$ are compared with
Gaussian polymer models.  Here, it is demonstrated that the effective diffusion 
coefficient required to approximately describe the radial autocorrelation
function brings the bounding SSS and WF approximations into agreement with
the simulated mean first contact time.

The parameters required to calculate the WF and SSS
approximations can be chosen in a couple of different ways.
For the lower bound, 
$\tau_{\scap{\mathrm{SSS}}}\deff$ calculated with either the simulated
potential of mean force 
$U(R) = -kT \log P_{\mathrm{eq}}(R)$ [Eq.(\ref{eq:tau_sss_quad})], 
or the harmonic potential $U(R) = 3R^2/2\langle R^2 \rangle$
[Eqs.(\ref{eq:sss_tau}--\ref{eq:x0})] 
with $\langle R^2 \rangle$
from simulation or the Rouse chain agree to within 10\% of each other for all
the sink radii considered.
This suggests that longitudinal ruggedness in the potential of mean force
does not significantly reduce the effective diffusion coefficient for 
this peptide;\cite{fn:jjp02_sssgauss}
this conclusion is also consistent with the analysis of quenching times for
the two shorter peptides presented by Yeh and Hummer.\cite{iy:gh:02}
For the upper bound  [Eq.(\ref{eq:wf_tau}) with the Gaussian correlations
Eqs.(\ref{eq:kk/k_wf}--\ref{eq:kk_wf_var})], 
$\tau_{\scap{\mathrm{WF}}}$ calculated 
with either $\phi(t)$ and $\langle R^2 \rangle$
from simulation or from the Rouse model differ by a few percent.
Although these calculated rates are in sufficient agreement for the
subsequent analysis, to emphasize independence of a Gaussian chain model, 
the reported mean first contact times use
the parameters directly from the simulation:
harmonic potential defined with 
$\langle R^2 \rangle_{\mathrm{sim}}$ for $\tau_{\scap{\mathrm{SSS}}}$, and 
$\phi_{\mathrm{sim}}(t)$ and $\langle R^2 \rangle_{\mathrm{sim}}$
for $\tau_{\scap{\mathrm{WF}}}$.

\begin{table}
\caption{\label{tbl:tau_cw3}
Diffusion limited mean first passage times of 
$\mathrm{CW}_3$ with different sink radii.}
\begin{ruledtabular}
\begin{tabular}{|c|cc|cc|cc|}
 &\multicolumn{2}{c|}{Simulation} 
 &\multicolumn{2}{c|}{$D_0 = 0.15\,\Asqps$}
 &\multicolumn{2}{c|}{$\bar{D}_0 = D_0/6$}\\
$a$[{\AA}] 
& $n_{\mathrm{int}}$\footnotemark[1]
& $\tau_{\mathrm{sim}}\, \mathrm{[ns]}$\footnotemark[2]
& $\tau_{\scap{\mathrm{SSS}}}\, \mathrm{[ns]}$\footnotemark[3]
& $\tau_{\scap{\mathrm{WF}}} \, \mathrm{[ns]}$\footnotemark[4]
& $\tau_{\scap{\mathrm{SSS}}}\, \mathrm{[ns]}$\footnotemark[5]
& $\tau_{\scap{\mathrm{WF}}} \, \mathrm{[ns]}$\footnotemark[6]\\ 
\hline
4.0& 816  & 2.8 $\pm$ 0.7&  0.27 & 0.67 & 1.6  & 4.0 \\
4.5& 901  & 2.5 $\pm$ 0.7&  0.23 & 0.61 & 1.4  & 3.7 \\
5.0& 870  & 2.1 $\pm$ 0.7&  0.20 & 0.56 & 1.2  & 3.4 \\
5.5& 1034 & 1.4 $\pm$ 0.3&  0.17 & 0.51 & 1.0  & 3.1 \\
6.0& 1297 & 1.3 $\pm$ 0.3&  0.15 & 0.46 & 0.90 & 2.8 \\
6.5& 1495 & 0.9 $\pm$ 0.2&  0.13 & 0.42 & 0.78 & 2.5 \\
\end{tabular}
 \end{ruledtabular}
\footnotetext[1]{
Number of contact events in simulation.
}
\footnotetext[2]{
Uncertainty is estimated as the standard error in the mean
found by the limiting block averaged error 
using the contact times of each configuration with $R(t) \ge a$ .
}
\footnotetext[3]{
Eqs.(\ref{eq:sss_tau}--\ref{eq:x0}) using the simulated mean
square distance,
$\langle R^2 \rangle_{\mathrm{sim}} = 140\,\A^2$, and $\deff = 2D_0$.
}
\footnotetext[4]{
Eq.(\ref{eq:wf_tau}) and Eqs.(\ref{eq:kk/k_wf}--\ref{eq:kk_wf_var}) using 
the simulated mean square distance,
$\langle R^2 \rangle_{\mathrm{sim}} = 140\,\A^2$, 
and pair correlation, $\phi_{\mathrm{sim}}(t)$.
}
\footnotetext[5]{
Same as [c], but with $\deff = 2\bar{D}_0$.
}
\footnotetext[6]{
Same as [d], but with time scaled as 
$t \rightarrow \bar{t} = (D_0/\bar{D}_0) t$,
i.e., using $\phi_{\mathrm{sim}}(\bar{t})$. 
}
\end{table}

As shown in Table \ref{tbl:tau_cw3}, the calculated SSS and WF approximations 
underestimate the simulated 
 mean first contact times for 
all sink radii considered.  The SSS approximation is calculated with 
$D_{\mathrm{eff}} = 2D_0$. If $D_0$ is treated as a fitting parameter
(similar to previous analysis of experimentally measured times),
the fit to $\tau_{\scap{\mathrm{SSS}}}$
defines a new effective monomer diffusion constant
$\bar{D}_0 = D_0/\alpha_{\scap{\mathrm{SSS}}}$
with $\alpha_{\scap{\mathrm{SSS}}} = \tau_{\mathrm{sim}}/\tau_{\scap{\mathrm{SSS}}}$.
From Table \ref{tbl:tau_cw3},
$\alpha_{\scap{\mathrm{SSS}}}$ ranges from approximately
7 for $a = 6.5\,\A$ to 10 for $a = 4.0\,\A$.
Following the same reasoning, the WF approximation also has an associated
effective diffusion coefficient.  
For overdamped motion, an effective diffusion constant
scales time as $\bar{t}\bar{D}_0 = t D_0$, so that using $D_0$  as a fitting parameter
gives $\bar{D}_0 = D_0/\alpha_{\scap{\mathrm{WF}}}$ with 
$\alpha_{\scap{\mathrm{WF}}} = \tau_{\mathrm{sim}}/\tau_{\scap{\mathrm{WF}}}$.
From Table \ref{tbl:tau_cw3}, 
$\alpha_{\scap{\mathrm{WF}}}$ ranges from 2 for $a = 6.5\,\A$ to 4 for $a = 4.0\,\A$.
(Note that $\alpha_{\scap{\mathrm{SSS}}} \ge \alpha_{\scap{\mathrm{WF}}}$ follows from
$\tau_{\scap{\mathrm{SSS}}} \le \tau_{\scap{\mathrm{WF}}}$.)
Since SSS and WF are upper and lower bound approximations, respectively,
the effective diffusion coefficient 
for the original chain should actually lie between the values obtained
from each approximation; 
then, $\tau_{\scap{\mathrm{SSS}}} \le \tau_{\mathrm{sim}} \le \tau_{\scap{\mathrm{WF}}}$.
For example, for the sink radius $a = 5.0\,\A$, the $\bar{D}_0$ should
satisfy $D_0/10 \alt \bar{D}_0 \alt D_0/4$.

\begin{figure}[htb]
\includegraphics[width=3.3in,clip=true]{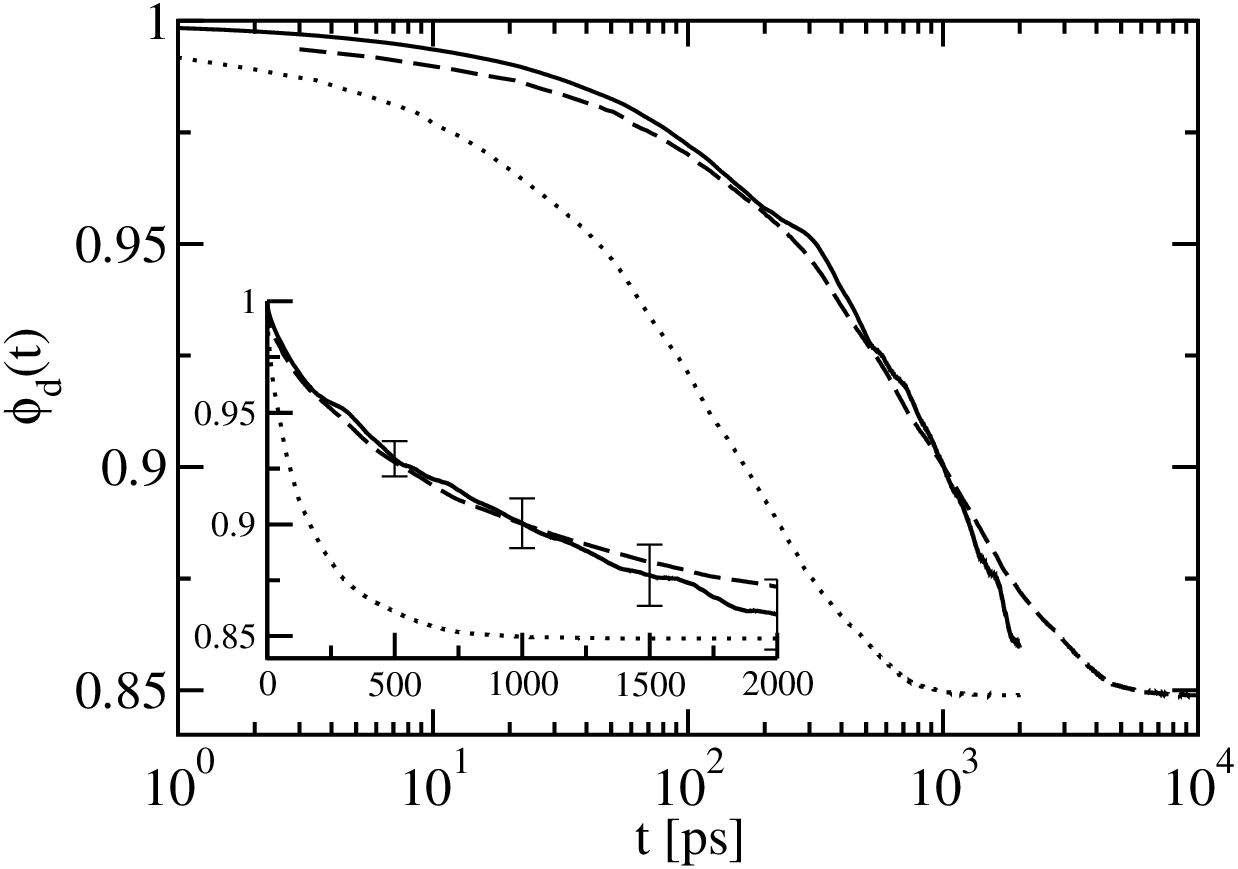}
  \caption{\label{fig:phi_d}
Distance pair correlation $\phi_{\mathrm{d}}(t)$ for $\mathrm{CW}_3$.
Solid: simulation; dotted: 
$\phi_{\mathrm{d}}^{\mathrm{g}}(t)$ using the simulated pair correlation
$\phi_{\mathrm{sim}}(t)$ in Eq.(\ref{eq:phi_d_g}); 
dashed: same as dotted curve with time
scaled $t \rightarrow \bar{t} = (D_0/\bar{D}_0) t$ with
$D_0/\bar{D}_0 = 6$.
Inset is same plot on linear time axis.
}
\end{figure}

\begin{figure}[htb]
\includegraphics[width=3.3in,clip=true]{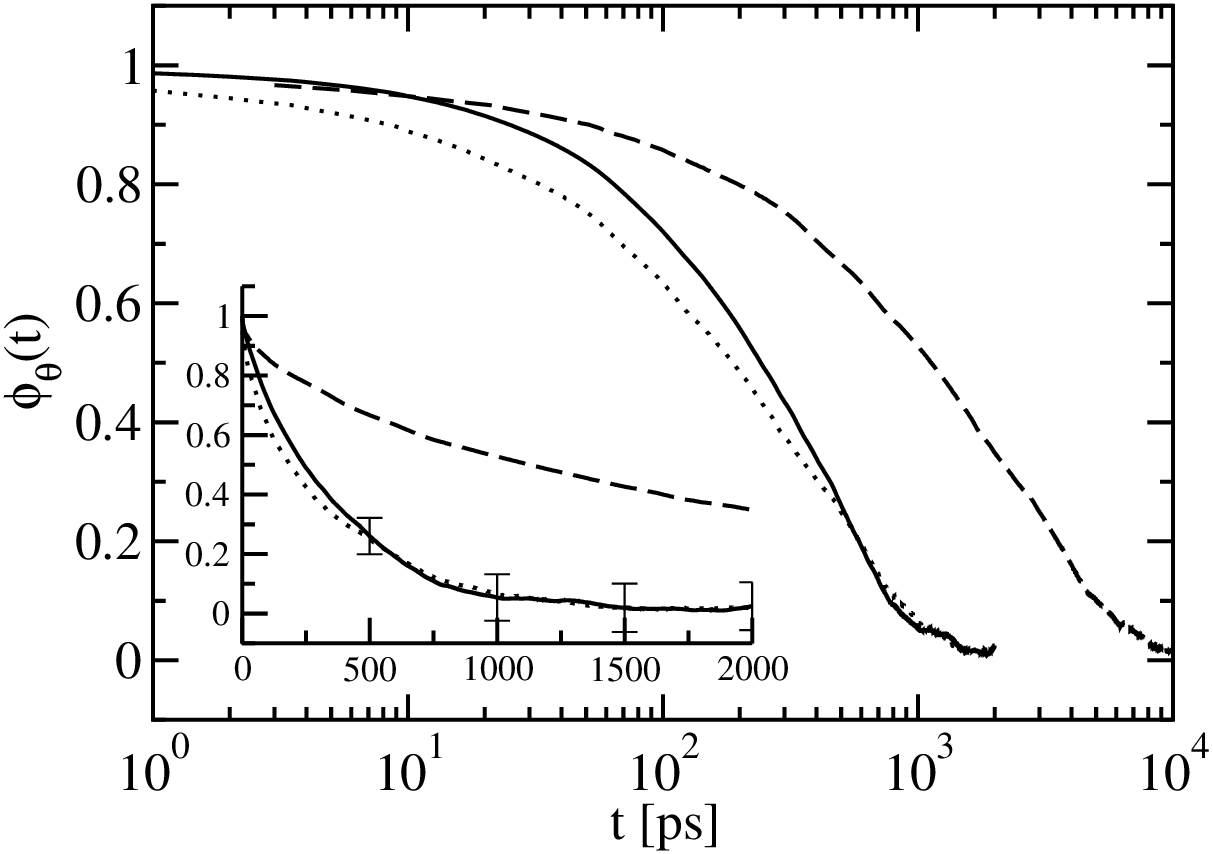}
  \caption{\label{fig:phi_th}
Angular pair correlation $\phi_{\theta}(t)$ for $\mathrm{CW}_3$.
Solid: simulation; dotted: 
$\phi_{\theta}^{\mathrm{g}}(t)$ using the simulated pair correlation
$\phi_{\mathrm{sim}}(t)$ in Eq.(\ref{eq:phi_th_g}); 
dashed: same as dotted curve with time
scaled $t \rightarrow \bar{t} = (D_0/\bar{D}_0) t$ with
$D_0/\bar{D}_0 = 6$.
Inset is same plot on linear time axis.
}
\end{figure}

Why do these approximations require a reduced effective diffusion 
coefficient to fit simulated contact rates?  
It is argued here that the effective diffusion coefficient
reflects the slow dynamics of the end-to-end distance.
In terms of comparing to a Gaussian model,
the small effective diffusion coefficient accounts for the discrepancy of 
using a Gaussian theory to describe the
non-Gaussian dynamics of the end-to-end vector.
The simulated correlations $\phi_{\mathrm{d}}(t)$ and $\phi_{\theta}(t)$
are shown as solid lines in Fig.~\ref{fig:phi_d} and Fig.~\ref{fig:phi_th}, 
respectively.  The dotted lines in these figures show the Gaussian expressions
$\phi_{\mathrm{d}}^{\mathrm{g}}(t)$ and $\phi_{\theta}^{\mathrm{g}}(t)$
[Eqs.(\ref{eq:phi_d_g}--\ref{eq:phi_th_g})] calculated from the simulated 
pair correlations $\phi_{\mathrm{sim}}(t)$.  
Comparing the simulated correlations with the 
Gaussian correlations tests the Gaussian statistics of $\bm{R}(t)$,
[i.e., this checks whether correlations can be accurately 
calculated with the isotropic Gaussian Green's function in Eq.(\ref{eq:G_R})].
Both the simulated radial and angular correlations decay more slowly
than the Gaussian model predicts.
However, 
the deviation of the angular correlation from Gaussian form is small
in comparison to the deviation of the radial correlations.
The semi-log plot in Fig.~\ref{fig:phi_d} shows that
the simulated $\phi_{\mathrm{d}}(t)$ and $\phi_{\mathrm{d}}^{\mathrm{g}}(t)$
have roughly the same shape, but are translated along the time axis, 
i.e., they differ by a constant scaling of time.
As shown by the dashed curve in Fig.~\ref{fig:phi_d},
$\phi_{\mathrm{d}}^{\mathrm{g}}(t)$ agrees reasonably well with the simulated
radial correlation if time scaled as 
$t \rightarrow \bar{t} = \alpha t$ with $\alpha = 6$.
For overdamped motion, this is equivalent
to an effective monomer diffusion coefficient $\bar{D}_0 = D_0/\alpha$.
As shown in Fig.~\ref{fig:phi} and Fig.~\ref{fig:phi_th},
this scaling applied to $\phi(t)$ and $\phi_{\theta}^{\mathrm{g}}(t)$
results in a poor agreement with the simulated correlations.
So, rescaling the diffusion constant to bring the
Gaussian and simulated radial correlation function into agreement,
makes $\phi(\bar{t})$ and $\phi_{\theta}^{\mathrm{g}}(\bar{t})$ 
disagree with the simulated dynamics.  

The dynamics of the end-to-end distance is the most relevant
to mean first contact times calculations.
Table \ref{tbl:tau_cw3} also shows the calculated times
using the effective monomer diffusion coefficient\cite{fn:jjp02_deff_scaled} 
($\bar{D}_0 = D_0/6$ with $\alpha = 6$)
determined from the fit to $\phi_{\mathrm{d}}(t)$.
With this scaling, 
the SSS and WF approximations bracket the simulated contact time, 
as they should since they are upper and lower bounds.  
Since the approximations provide bounds 
for a rather broad range of $\alpha$, 
these results offer only speculative support that the 
effective diffusion coefficient determined by the mean first contact
time reflects the dynamics of $R(t)$.
Because attention to $\phi_{\mathrm{d}}(t)$ arises quite naturally from
general considerations, the consistency of the analysis is very suggestive.

\section*{Conclusion and Outlook}

The simulated diffusion limited mean first contact time agrees with
the bounds provided by the WF and SSS approximations for a Gaussian
chain if the diffusion coefficient is reduced.
When viewed as completely independent approximations, it would seem that
the SSS and WF approximations would each require different rescalings.
It was shown, however, that 
the two approximations are complementary variational bounds, giving a range 
of values for the scaling because the approximations
should merely bracket the measured times,
not necessarily fit them each independently.

It was also demonstrated that the 
relative end-to-end distance of the simulated peptide 
relaxes more slowly than the relaxation from Gaussian dynamics.
It is argued that the effective diffusion coefficient appropriate 
for the calculated mean first contact time also
describes the end-to-end distance dynamics.
Evidence supporting this connection is weakened by the broad interval between
the bounds provided by the SSS and WF approximations.
Tighter bounds (and other approximations) 
can be generated by the variational formalism
through more flexible trial functions, helping to make support for this
relationship more precise.
If this relationship is indeed valid,
one can focus on $\phi_{\mathrm{d}}(t)$
rather than the mean first contact time itself to improve the model.
Furthermore, 
the analysis demonstrates that accurately modeling $\phi(t)$ 
is not the complete description of the polymer dynamics, 
for the Gaussian statistics are checked
using $\phi_{\mathrm{sim}}(t)$ rather the pair correlation 
from a polymer model.

It should be noted that Gaussian dynamics and the 
reduction of a multi-bead model
to a single variable model are separate issues.  The variational
bounds $\tau_{\scap{\mathrm{SSS}}} \le \tau \le \tau_{\scap{\mathrm{WF}}}$ 
is a relationship between
the mean contact time of a single variable model (with a specific $\deff$)
to the corresponding multi-bead model.  
Reducing the description to a single variable model
provides the microscopic interpretation of $\deff$ as the relative free
diffusion coefficient of the end monomers.
In general, both models may be non-Gaussian,
with the potential in the reduced model replaced by the potential of mean 
force of the chain.
It seems that a microscopic interpretation of the slow end-to-end distance 
relaxation would require a multi-bead description, perhaps
leading to an effective single variable description which is easier to handle
analytically.  
In any case, the comparison of the end-to-end distance and 
angular relaxation should guide the development of a microscopic model
that extends beyond the simple Gaussian dynamics of a single reaction coordinate.

In a recent paper, Lapidus \emph{et al.}\cite{lsesh:02} present
further experimental work and theoretical modeling of quenching 
in $\mathrm{CW}_k$ peptides.
In these new experiments, the viscosity dependence of the quenching rate in
$\mathrm{CW}_k$ peptides is measured, 
isolating the diffusion limited component of the measured quenching rate.
Fitting the measured diffusion limited quenching rate with a single
variable approximation corresponding to a stiff chain
gives  $\bar{D}_0 \approx D_0/10$.  
This is the same factor obtained
from the fit to the simulations presented here with a reaction 
radius of $a \le 5 \, \A$ had the SSS approximation been used to 
define the effective monomer 
diffusion coefficient $\bar{D}_0$ (see Table~\ref{tbl:tau_cw3}).
To avoid confusion, it is emphasized again
that the value found from our analysis of the simulation,
$\bar{D}_0 = D_0/6$, 
was determined by the dynamics of the end-to-end distance,
not the lower bound approximation to the contact time.

In the theoretical modeling portion of Lapidus \emph{et al.},\cite{lsesh:02}
the mean first contact time of the end-to-end monomers is studied through
Langevin simulations of a model that includes
excluded volume and chain stiffness (provided by the dihedral angle potential),
but no other non-local interactions between the monomers.
The dynamics of the squared end-to-end distance from the simulation 
is analyzed in some detail with respect to the single reaction coordinate
description.
While the relationship between end-to-end distance dynamics
and other correlation functions (angular correlations, for example) is
not considered,
Lapidus \emph{et al.}\cite{lsesh:02} do emphasize
the connection between the effective diffusion coefficient
appropriate for the mean quenching time and the relaxation the 
end-to-end distance.
Comparing to experimental data, their analysis suggests that the 
effective diffusion constant determined by a fit to the single variable
approximation is not explained by the local properties of the backbone, 
but must involve interactions left out of the Langevin simulations; 
e.g., non-local contacts or explicit solvent.  
This very interesting conclusion limits the possible mechanism responsible 
for the slow non-Gaussian relaxation of the end-to-end distance.

The elements missing from these Langevin simulations
are included in the all-atom molecular dynamics simulations 
analyzed in the present paper.  
Here, an improved model with gated diffusion seems promising:
the trajectory of R(t) shows relatively long waiting times separating
periods of larger diffusional flights in distance 
reminiscent of continuous time random walks, for example.
To be consistent with the simulated dynamics,
such a model should have the property that the gating retards
the end-to-end distance relaxation, but has a relatively smaller effect
on the angular relaxation, at least for chains not too long compared to 
the persistence length.  
More refined analysis of the conformational transitions in the 
simulation should give insight into how to formulate a gated diffusion
model; this is left for future work.
Any microscopic model of the gated diffusion
should be constrained by the reasonable expectation that the 
dynamics become Gaussian in the limit of very long chains.

\begin{acknowledgments}
I thank William Eaton making Ref.~\onlinecite{lsesh:02} available
before publication.  I also thank Angel Garcia for 
guidance in carrying out the simulations and helpful suggestions;
and Peter Wolynes for our many discussions on this topic 
and his insightful comments on the manuscript.
This work was supported by a Director's Postdoctoral Fellowship and 
the Los Alamos National Laboratory Directed Research Fund (Ben McMahon, PI).
\end{acknowledgments}

\section*{Appendix: 
$\phi(t)$ and the short time $\deff$ for Gaussian chains} \label{ap:phi_deff}

This appendix gives the pair correlations $\phi(t)$ and derives
the single variable approximation effective diffusion coefficient
$\deff$ for a general Gaussian chain.  
The effective diffusion coefficient, derived 
here as a short time approximation,
can also be obtained from the variational lower bound [Eq.(\ref{eq:deff})] 
as well as a ``local equilibrium'' approximation.\cite{rwp:rz:as:96}

First, the diffusion matrix in the Smoluchowski equation 
[Eq.(\ref{eq:smol_chain})] is written as a product of a diffusion constant 
and a unitless matrix $D_{ij} = D_0 \tilde{D}_{ij}$.
The Smoluchowski equation can be diagonalized by transforming
into normal modes
$\bm{r}_i = \sum_p Q_{ip}\tilde{\bm{r}}_p$ where the transformation
matrix $\bm{\mathsf{Q}}$ is defined by 
\begin{eqnarray}\label{eq:modes}
\bm{\mathsf{Q}}^{-1}\bm{\mathsf{\tilde{D}\Gamma Q}}
       &=&\rm{diag}(\lambda_p) \nonumber\\*
\bm{\mathsf{Q}}^{-1}\bm{\mathsf{\tilde{D}}}\bm{\mathsf{Q}}^{-1T}
       &=&\rm{diag}(\nu_p) \\*
\bm{\mathsf{Q}}^{T}\bm{\mathsf{\Gamma Q}}
       &=&\rm{diag}(\mu_p) \nonumber
\end{eqnarray}
with $\lambda_p = \nu_p \mu_p$, $p = (0 \ldots N-1)$.  
The pair correlations of the end to end
vector, $\bm{R} = \bm{r}_N - \bm{r}_1$, can then be written in terms
of the normal modes
\begin{equation}\label{eq:phi_modes}
\phi(t) = 
\frac{b^2}{\left\langle R^2 \right\rangle}
\sum_{p \ne 0}\frac{1}{\mu_p} (Q_{Np} - Q_{1p})^2
\exp \left[- \frac{3D_0}{b^2}\lambda_p t \right].
\end{equation}

Next, consider the effective diffusion constant of
the single variable approximation to the Gaussian chain.
The single variable approximation (such as SSS) treats $\bm{R}(t)$
as the only dynamical variable.  The Green's function for
$\bm{R}(t)$ [Eq.(\ref{eq:G_R})] satisfies the
diffusion equation 
\begin{equation}\label{eq:smol_sss}
\partial_t G(\bm{R}t|\bm{R}_0) = {\cal D}(\bm{R};t) G(\bm{R}t|\bm{R}_0),
\end{equation}
where ${\cal D}$ is the single variable diffusion operator
\begin{equation}\label{eq:diff_op_R}
{\cal D}(\bm{R};t) = 
D(t)
\frac{\partial}{\partial \bm{R}} P_{\mathrm{eq}}(\bm{R})
          \cdot
          \frac{\partial}{\partial \bm{R}} \frac{1}{P_{\mathrm{eq}}(\bm{R})}
\end{equation}
with a time dependent diffusion coefficient 
\begin{equation}\label{eq:dt_sss}
D(t) = -\frac{\left\langle R^2 \right\rangle}{3} 
\frac{\mathrm{d}}{\mathrm{d}t} \log \phi(t).
\end{equation}
This equation is exact.  
The single variable approximation requires a choice for an
effective diffusion constant $\deff$ to replace
$D(t)$.  
Bicout and Szabo\cite{djb:as:98b} discuss the choice of 
$\deff$ and demonstrate some limitations of the single 
variable approximation in the context of rate calculations 
of electron transfer in non-Debye solvents.

The expression for $\deff$ derived from the variational
lower bound [Eq.(\ref{eq:deff})] is equivalent
to a short time approximation, $\deff = D(0)$.
Using $\lambda_p = \nu_p \mu_p$ and
$\tilde{D}_{ij} = \sum_p \nu_p Q_{ip}Q_{jp}$ from Eq.(\ref{eq:modes}),
$\dot{\phi}(0)$ reduces to
\begin{eqnarray}\label{eq:phidot0_1}
\dot{\phi}(0)
&=& 
- \frac{3D_0}{\left\langle R^2 \right\rangle}
 \sum_{p \ne 0}\frac{\lambda_p}{\mu_p} (Q_{Np} - Q_{1p})^2 \\*
&=&
- \frac{3D_0}{\left\langle R^2 \right\rangle}
( \tilde{D}_{NN} + \tilde{D}_{11} - \tilde{D}_{1N} - \tilde{D}_{N1} ) 
.
\end{eqnarray}
Since $\phi(0) = 1$, $\deff = D(0)$ becomes
\begin{equation}\label{eq:deff_short_2}
\deff = D_{NN} + D_{11} - D_{1N} - D_{N1}
\end{equation}
which is Eq.(\ref{eq:deff}).

As a specific example, the Rouse chain (nearest neighbor connectivity)
and uniform diagonal friction, $D_{ij} = D_0\delta_{ij}$ has
modes and relaxation rates
\begin{eqnarray}\label{eq:rouse_modes}
Q_{ip} &=& \sqrt{\frac{2 - \delta_{p0}}{N}}
           \cos \left(\frac{p(i-1/2)}{\pi N} \right) \\*
\lambda_p &=& \mu_p = 4 \sin^2\left(\frac{p\pi}{2N}\right)\nonumber,
\end{eqnarray}
and the mean square end-to-end distance is
\begin{equation}\label{eq:rouse_r2}
\left\langle R^2 \right\rangle = (N - 1) b^2.
\end{equation}
The effective short time diffusion constant,
\begin{equation}\label{eq:rouse_deff}
\deff = 2D_0,
\end{equation}
is the relative diffusion constant of the chain ends.
Internal friction (of the form in Ref.~\onlinecite{cwm:mcw:85})
reduces $\deff$.

In the single variable approximation to the end-to-end chain dynamics,
the pair correlation decays with a single time constant
\begin{equation}
\phi(t) = \exp \left[ -\frac{3 \deff}{\left\langle R^2 \right\rangle} t\right].
\end{equation}



\begin{thebibliography}{55}
\bibitem[{fn:({\natexlab{a}})}]{fn:jjp02_sokolov}
\bibinfo{note}{Very recently,\cite{sokolov:02} this problem has be reduced to
  the the solution of an integral equation for the special case when the
  quenching has a delta function distance dependence.}
\bibitem[{fn:({\natexlab{b}})}]{fn:jjp02_kupper}
\bibinfo{note}{This form of the upper bound is given in
  Ref.~\onlinecite{jjp:pgw:99} Eq.(B9) with $g = \xi$.}
\bibitem[{fn:({\natexlab{c}})}]{fn:jjp02_1dvar}
\bibinfo{note}{The variational bounds can be similarly developed for the single
  variable equation $\left[{\cal L}(\bm{R}) + \epsilon k \right] \hat{\rho} =
  1$. For finite $\epsilon$, bounds of this one-dimensional problem generated
  from different trial functions has been addressed in
  Ref.~\onlinecite{jjp:pgw:99}. We do not pursue this approach for
  Eq.(\ref{eq:sss_eq}) because the diffusion limited rate can be calculated
  exactly through Eq.(\ref{eq:tau_sss_quad}).}
\bibitem[{fn:({\natexlab{d}})}]{fn:jjp02_sssgauss}
\bibinfo{note}{The agreement, however, is to some degree accidental since the
  SSS approximation depends in part on the weight of $P_{\mathrm{eq}}(R)$ at
  small distances, where the Gaussian is a poor approximation to the simulated
  potential of mean force. The additional weight of the
  Gaussian at small $R$ is offset to some extent by the peak centered around
  $4\, \A$ in the simulated $P_{\mathrm{eq}}(R)$.}
\end{thebibliography}

\newpage
\printtables
\newpage
\printfigures

\end{document}